\documentclass{egpubl}

\usepackage{subfig,enumitem}

 \JournalSubmission % uncomment for submission to Computer Graphics Forum
\usepackage[T1]{fontenc}
\usepackage{dfadobe} 

\usepackage{cite} % comment out for biblatex with backend=biber
\BibtexOrBiblatex
\electronicVersion
\PrintedOrElectronic
\ifpdf \usepackage[pdftex]{graphicx} \pdfcompresslevel=9
\else \usepackage[dvips]{graphicx} \fi

\usepackage{egweblnk}

\usepackage{xcolor}
\definecolor{changescolor}{RGB}{0, 63, 191}
\usepackage[final,commandnameprefix=ifneeded]{changes}

\title[GeoDEN: A Visual \added{Exploration }\deleted{Analytics }Tool for \added{Analyzing }\deleted{Exploring }the Geographic Spread of Dengue Serotypes]%
  {GeoDEN: A Visual \added{Exploration }\deleted{Analytics }Tool for \added{Analyzing }\deleted{Exploring }the Geographic Spread of Dengue Serotypes}

\author[A. Marler et al.]
{\parbox{\textwidth}{\centering 
A. Marler$^{1}$\orcid{0009-0000-2604-0201},
Y. Roell$^{2}$\orcid{0000-0002-1146-6047}, 
S. Knoblauch$^{3}$\orcid{0000-0003-3077-8094},
J.P. Messina$^{4}$\orcid{0000-0001-7829-1272}, 
T. Jaenisch$^{2,5}$\orcid{0000-0001-8252-7444}, 
M. Karimzadeh$^{1}$\orcid{0000-0002-6498-1763}
  }
  \\
{\parbox{\textwidth}{\centering 
   $^1$University of Colorado Boulder, $^2$Colorado School of Public Health, Center for Global Health, $^3$Heidelberg University, $^4$University of Oxford, $^5$Heidelberg Institute of Global Health (HIGH), Heidelberg University Hospital%#69120 Heidelberg, Germany% Boulder, Colorado, United States  
  }
}
}
\begin{document}
\teaser{
 \includegraphics[width=1\linewidth]{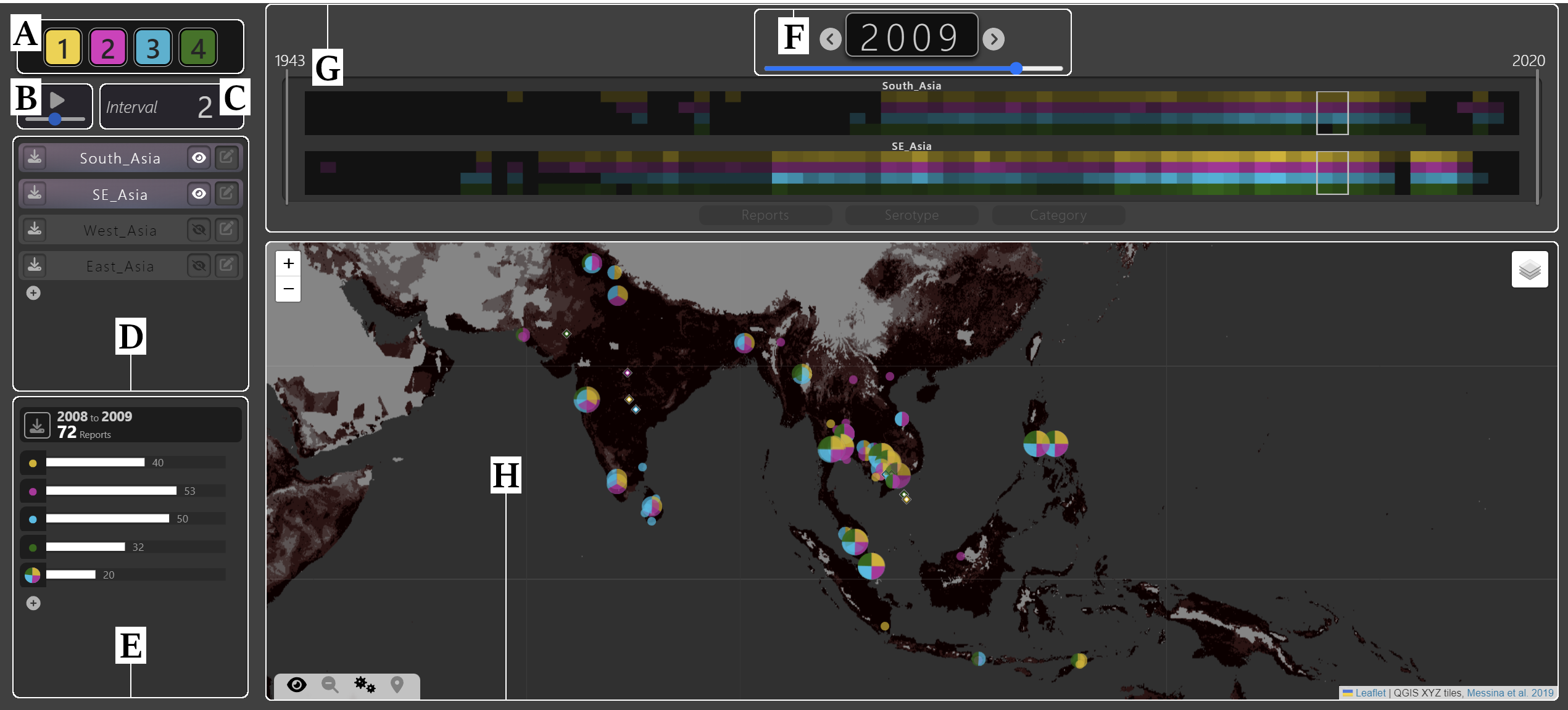}
 \centering
 \caption{\textbf{A,} Serotype Selection: choose active serotypes. \textbf{B,} Animation Controls: toggle and adjust animation speed. \textbf{C,} Interval Length Editor: modify number of years visualised. \textbf{D,} Region Faceting: define regions for visualisation. \textbf{E,} Co-Occurrence Histogram: examine frequency of serotype co-occurrence. \textbf{F,} Year Editor: adjust the year. \textbf{G,} Heatmap Timeline: compare regions and serotypes over time. \textbf{H,} Map: view reports, centroids, trajectories, and environmental suitability. }
\label{fig:teaser}
}

\maketitle
%-------------------------------------------------------------------------
\begin{abstract}
  Static maps and animations remain popular in spatial epidemiology of dengue, limiting the analytical depth and scope of visualisations. Over half of the global population live in dengue endemic regions. Understanding the spatiotemporal dynamics of the four closely related dengue serotypes, and their immunological interactions, remains a challenge at a global scale. To facilitate this understanding, we worked with dengue epidemiologists in a user-centered design framework to create GeoDEN, an exploratory \replaced{visualisation }{visual analytics }tool that empowers experts to investigate spatiotemporal patterns in dengue serotype reports. The tool has several linked visualisations and filtering mechanisms, enabling analysis at a range of spatial and temporal scales. To identify successes and failures, we present both insight-based and value-driven evaluations. Our domain experts found GeoDEN valuable, verifying existing hypotheses and uncovering novel insights that warrant further investigation by the epidemiology community. The developed visual \added{exploration }\deleted{analytics }approach can be adapted for exploring other epidemiology and disease incident datasets.
%-------------------------------------------------------------------------
% ACM CCS 1998
% (see https://www.acm.org/publications/computing-classification-system/1998)
% \begin{classification} % according to https://www.acm.org/publications/computing-classification-system/1998
% \CCScat{Computer Graphics}{I.3.3}{Picture/Image Generation}{Line and curve generation}
% \end{classification}
%-------------------------------------------------------------------------
% ACM CCS 2012
% (see https://www.acm.org/publications/class-2012)
%The tool at \url{http://dl.acm.org/ccs.cfm} can be used to generate
% CCS codes.
%Example:
\begin{CCSXML}
<ccs2012>
 <concept>
  <concept_id>10003120.10003145.10003147.10010365</concept_id>
  <concept_desc>Human-centered computing~Visual analytics</concept_desc>
  <concept_significance>300</concept_significance>
  </concept>
 <concept>
  <concept_id>10003120.10003145.10003147.10010887</concept_id>
  <concept_desc>Human-centered computing~Geographic visualisation</concept_desc>
  <concept_significance>300</concept_significance>
  </concept>
 <concept>
  <concept_id>10003120.10003123.10010860.10010859</concept_id>
  <concept_desc>Human-centered computing~User centered design</concept_desc>
  <concept_significance>300</concept_significance>
  </concept>
 </ccs2012>
\end{CCSXML}

\ccsdesc[300]{Human-centered computing~Visual analytics}
\ccsdesc[300]{Human-centered computing~Geographic visualisation}
\ccsdesc[300]{Human-centered computing~User centered design}

\printccsdesc 
\end{abstract} 
%-------------------------------------------------------------------------
\section{Introduction}

With the increasing availability of global disease datasets, Visual Analytics (VA) is gaining further recognition and use in spatial epidemiology \cite{chishtie_visual_2020}, especially in the exploration of serotypes and their interactions in diseases like dengue. Dengue, caused by serotypes DENV1-DENV4, presents a global health threat, exacerbated by factors like urbanization and climate change. While previous studies and the common domain practice have focused on static mapping or animation of the incidence of these serotypes \cite{preim_survey_2020}, there remains a gap in the utilization of VA tools for dynamic and interactive analysis of spatiotemporal patterns in complex datasets effectively. The potential of VA in epidemiology lies in its ability to provide rapid insights over multiple dimensions (time, space, and attributes), facilitating a deeper understanding of serotype distribution, interaction, and evolution over time and space. 

In this paper, we present GeoDEN, an exploratory visual analytics tool to support epidemiological investigation into the spread and interaction of dengue serotypes at a global scale. GeoDEN helps experts develop a better understanding of global vulnerability to dengue, population level transmission, and serotype interactions over time and space.  In epidemiology, various disease incidence datasets are collected for research and intervention \cite{nakakana_validation_2020}. Through computation\replaced{-}{ }assisted visualisation of global dengue serotype occurrence, our tool enables the generation of new hypotheses, as well as facilitating the testing and validation of ideas postulated in the literature. Insights generated using GeoDEN can be integrated into future research, supporting and directing vaccine development, aid, and intervention programs.

In addition to an insight-based evaluation, we conduct a value-driven evaluation, offering distinct and complementary perspectives of the tool.
Our contributions in this paper includes the iterative design, implementation and evaluation of an exploratory visual analytics system for a common problem in epidemiology. We document and analyse the successes and failures of our approach to guide future research in visual analytics and epidemiology.

%-------------------------------------------------------------------------
\section{Related Work}

The immunological interaction between subsequent dengue infections, and its association with a severe course of disease, was first described in the 1980s in Thailand \cite{simmons_current_2012,cummings_travelling_2004} and was later known as ‘antibody dependent enhancement’ of infection. The study of underlying biological mechanisms revealed that the sequence and time interval between infections play a crucial role in shaping the risk of severe disease \cite{ohainle_dynamics_2011}. While distinct serotype movement patterns have been described in certain areas \cite{cummings_travelling_2004, ohainle_dynamics_2011,priyamvada_humoral_2017,gainor_uncovering_2022,houghton-trivino_dengue-yellow_2008}, tools to analyse serotype movements on a global or regional scale–which is the focus of this paper–are lacking.

Interactive analysis of the global spread of DENV serotype enhances our understanding of dengue spread, endemicity, and disease severity. Regions with multiple serotypes are more likely to produce severe disease \cite{wilder-smith_epidemiology_2013}. As pan-serotypic vaccines are not yet available, vaccine development strategies and eventual distribution currently depend on understanding which serotypes are present in which regions.

Monitoring regional serotype switches helps in assessing the risk of potential outbreaks. Since each serotype can cause a unique immune response in individuals, the introduction of a new serotype in an area with existing immunity to other serotypes can lead to more severe forms of the disease, as seen in cases of secondary infections. Larger DENV outbreaks tend to co-occur with serotype switches \cite{lee_dengue_2012}.

Data visualisation in epidemiology with a focus on disease spread and transmission often feature spatial and temporal components \cite{clements_spatial_2023,hadfield_nextstrain_2018,yi_pictureeaedes_2023,arora_serotracker_2021,dykes_visualization_2022,chui_visual_2011,preim_survey_2020}. Maps are used ubiquitously, a trend reinforced by the increasing prevalence of geospatial analytics in public health and epidemiology \cite{greenough_beyond_2019}. Spatiotemporal data is often represented as spatial data faceted by time or temporal data faceted by space \cite{messina_global_2014,clements_spatial_2023,hadfield_nextstrain_2018,yi_pictureeaedes_2023,dykes_visualization_2022,chui_visual_2011,andrienko_space_2010,wei_global_2017,andrienko_exploratory_2006}. Alternatively, spatiotemporal aspects can be integrated, such as overlaying temporal information onto geographic visualisation. For instance, geographic heatmaps can represent time spent in a location, or a line that shows movement over time \cite{andrienko_space_2010,castro_spatiotemporal_2021,andrienko_visual_2013}. Spatial aspects can also be visualised in a temporal visualisation where an axis represents time while another variable, like colour in a stacked bar chart, represents space \cite{brito_lying_2021}.

Temporal data can be shown in heatmaps, line charts, barcharts, streamgraphs, and horizon graphs \cite{clements_spatial_2023,chui_visual_2011,wei_global_2017,castro_spatiotemporal_2021}, providing researchers multiple options for user-friendly temporal visualisations. Animated maps, another method of spatiotemporal visualisation, are more effective with user interaction \cite{fabrikant_visual_2008,dibiase_animation_1992}.

Visual analytics (VA) has become increasingly popular in epidemiology as available data has increased \cite{chui_visual_2011,preim_survey_2020}, and is often used to enhance understanding of complex datasets. In epidemiology, VA enables an interactive environment to support exploration and analysis of disease evolution and movement \cite{clements_spatial_2023,hadfield_nextstrain_2018,yi_pictureeaedes_2023,arora_serotracker_2021,dykes_visualization_2022,chui_visual_2011,rydow_development_2022,preim_survey_2020}. Developing VA tools requires interdisciplinary collaboration, drawing insights from both domain experts and visualisation researchers \cite{sedlmair_design_2012,chui_visual_2011,preim_survey_2020,andrienko_space_2010,meyer_criteria_2019,lloyd_human-centered_2011,mccurdy_action_2016}. 

Several visualisation tools have been created to monitor disease spread, especially for the purpose of tracking genetic changes \cite{wood_designing_2010,carroll_visualization_2014,chishtie_visual_2020}. DENV evolution and spread is facilitated by NextStrain. While occurrences for the four serotypes can be observed over time, it was created with the primary goal of visualising the serotype evolution in real time; it does not support the analysis of dengue on regional scales, the interaction of serotypes, or the modeling and visualisation of disease movement over time (referred to as trajectory summarization, where a trajectory is the path of a disease over time) \cite{hadfield_nextstrain_2018}. Another VA system, PICTUREE was created for the purpose of predicting future outbreaks of dengue. PICTUREE uses machine learning models with data on DENV occurrence, mosquito occurrence, google trends, and weather data. While it is a sophisticated system, it was not created to deeply explore historical reports datasets and serotype interactions, rather to predict future cases \cite{yi_pictureeaedes_2023}. GeoDEN fills a gap: it enables the exploration and analysis of historical DENV reports at multiple spatiotemporal scales to support the investigation of DENV serotype movements and spread on regional and global scales. VA tools are increasingly created for use in epidemiology, however, none have the specifications required by our domain-expert users, who need a tool uniquely created for historical DENV serotype reports \cite{carroll_visualization_2014}.
%-------------------------------------------------------------------------
\section{Design}

Our design process for GeoDEN utilized an iterative user-centered design process to solve a domain specific problem.

%-------------------------------------------------------------------------
\subsection{Iterative User-Centered Design}

GeoDEN was developed in consultation with three experts in arboviruses, dengue epidemiology, and clinical practice—all co-authors of this article and belonging to the target-user group of GeoDEN.
It is worth noting that co-authorship of this visualisation paper was not their initial goal, neither promised to these experts. Rather, they wanted a visualisation tool to help them with their research. It was only after implementation of the tool and conducting the evaluation (Section 5) that we included these domain experts as co-authors, given their involvement in the co-design of GeoDEN. Initially, they sought an animated map of a global dengue serotype reports dataset \cite{messina_global_2014} to better understand how serotypes co-occur, move, and compete over time given the intricate interaction between population immunity and susceptibility to new serotypes. Through 18 informal formative evaluations over 7 months with intermediate mockups and prototypes, eight design requirements were identified. To fulfill these requirements, a comprehensive VA tool–called GeoDEN–was implemented, with informal formative evaluations and updated versions.

It is important to note that each expert had a different level of involvement in the co-design of GeoDEN. One guided the design throughout the process. The two other experts only provided feedback for improvement at the later stages of implementation. Specifically, Dr. Jaenisch (referred to as expert 1 in the evaluations) brought the domain issue and dataset to us; he was therefore the most involved, present for all 18 informal evaluations. From these early discussions, we abstracted the primary tasks and dataset characteristics, which led to the identification of design requirements 1, 2, 3, 4, 6, and 7. After the development of a functional prototype, we introduced the project to Dr. Messina (expert 2), the creator of the dataset. Dr. Messina provided valuable feedback at this stage through formative evaluation, refining implementations for design requirements 5 and 8. Finally, after the system was in alpha version, Mr. Knoblauch (expert 3) provided feedback, which informed modifications to our solutions to the design requirements. All three experts then participated in formal evaluations on real-world data, described in Section 5.

%-------------------------------------------------------------------------
\subsection{Domain Characterization and Abstraction}

GeoDEN’s design was driven by a domain-specific problem provided by expert epidemiologists focusing on dengue and other arboviruses. Dengue viruses (serotypes DENV1-DENV4), transmitted by \textit{Aedes} mosquitoes, impact urban and peri-urban areas in tropical and subtropical regions worldwide. The ranges of transmission are expanding due to global trends such as urbanization and climate change. The global incidence of dengue was reported at a record high in 2024; countries in the Americas alone reported 9.7 million cases in the first \replaced{six}{6} months of 2024, which is more than twice the number of cases reported in 2023\cite{noauthor_increased_2024}. Dengue has been estimated to affect around 100 million people worldwide with symptomatic disease, although there are likely at least 400 million actual cases per year as the majority are asymptomatic or manifest with minor flu-like illness \cite{bhatt_global_2013, zeng_global_2021}. Infection with one of the four serotypes is thought to provide long-term immunity for that serotype, however, following infections with heterologous serotypes are more likely to result in severe disease \cite{simmons_current_2012}. As such, it is crucial to understand the global distribution and overlap of each DENV serotype, as well as their movements across time, regions, and populations. Disease caused by infection with subsequent serotypes also complicates vaccine development and deployment strategies \cite{izmirly_challenges_2020}, making it especially important to elucidate serotype spread through patterns of global and regional movement\cite{liebman_spatial_2012}.

DENV serotype occurrence information was compiled in a study which evaluated 10,529 case reports of dengue. Through filtering reports with insufficient metadata, reports with low confidence, and duplicates that report the same cases, a final dataset of 3,260 reports was generated \cite{messina_global_2014, messina_global_2014-1}. The study presented a series of static maps showing the worldwide expansion of the four serotypes, conveying the expansion of disease establishment and hyperendemicity in many parts of the world \cite{messina_global_2014, messina_global_2014-1}. The original spatial and temporal summary visualisations did not provide the ability to explore for deeper insights. An interface to enable the interaction and interrogation of dengue serotype reports by an epidemiologist would facilitate the discovery of understudied patterns at a variety of spatial and temporal scales. The ability to select, facet, and animate through the data could reveal events that may only occur over a few years or between a few countries. Our domain experts recounted their challenges with exploring this large spatiotemporal dataset spanning multiple decades and requested a visual analytics solution.

The dataset includes the attributes of latitude, longitude, country, year, and dengue serotype presence. Serotype presence is reflected though four categorical attributes, holding if each serotype was reported, and the ordinal attribute of total serotypes present. The dataset suffers from spatial sparsity, temporal inconsistencies, and biases due to unequal reporting infrastructure in certain regions. Conversations with the domain experts made it clear that these challenges are common with these epidemiological datasets.

Through iterative discussions with three domain experts, we identified core user tasks to overcome the above challenge:
\begin{itemize}
    \item \textbf{T1: Identifying serotype co-occurrences.} Experts needed to uncover patterns of serotype overlap, essential for understanding severe disease risks driven by immunological interactions.
    \item \textbf{T2: Visualizing serotype movement over time.} Tracking the movement of serotypes at regional and global scales was necessary to hypothesize transmission pathways and assess intervention strategies.
    \item \textbf{T3: Comparing reports at broader scales.} Experts required the ability to analyze serotype distributions across user-specified spatial and temporal contexts.
\end{itemize}

Considering Brehmer and Munzner's multi-level typology of visualization tasks\cite{brehmer_multi-level_2013}, T1 involved discovering, exploring, and identifying patterns using visual encodings, selection, navigation, filtering, and aggregation of reports with multiple serotypes reported. T2 entailed discovering, exploring, and identifying through visual encodings, navigation, filtering, and aggregation of items which are possibly correlated. Lastly, T3 required discovering, browsing, exploring, comparing, and summarizing items through visual encodings, selection, navigation, filtering, and aggregation of reports within specified spatial and temporal contexts. To support these three tasks, we developed the following eight design requirements (DR).

%-------------------------------------------------------------------------
\subsection{ Design Requirements }

\begin{enumerate}[label=\textbf{DR\arabic*}, left=0pt, labelwidth=3em, labelsep=1em]
  \item Visualise reports of dengue serotypes in space and time.
  \item Provide spatial selection and filtering based on user-defined lists of countries or regions, with instant overview and details on demand. 
  \item Provide temporal selection and filtering based on user-defined time periods, with instant overview and details on demand.
  \item Summarize the spatial distribution of serotypes for selected regions, time periods, and serotypes. 
  \item Summarize disease trajectories and spatiotemporal movements for selected regions, time periods, and serotypes. 
  \item Enable the interactive animation of all components, including reports, trajectories, and spatial distribution summaries.
  \item Enable the comparison of spatiotemporal co-occurrence and the inference of interactions between user-defined subsets of serotypes for selected regions and time periods.
  \item Enable comparison of \added{observed spatial patterns (}serotype reports, trajectories and spatial distribution summaries\added{)} \deleted{(observed spatial patterns) }against\added{ expected spatial pattern} \added{(a map showing where dengue is predicted to occur given environmental conditions, referred to as an environmental suitability map).}\deleted{an environmental suitability map for dengue (expected spatial pattern).}
\end{enumerate}

\noindent The design requirements \textbf{DR1-DR8} were driven by tasks \textbf{T1-T3}:

\begin{itemize}

    \item  \textbf{T1}, the goal of analyzing co-occurrences, informed \textbf{DR7} (compare spatiotemporal co-occurrences and interactions).
    
    \item  \textbf{T2}, the need to visualize serotype movements, led to \textbf{DR5} (summarize disease trajectories) and \textbf{DR6} (enable interactive animations).
    
    \item  \textbf{T3}, the requirement to compare reports at broader scales, guided \textbf{DR2} (spatial selection and filtering), \textbf{DR3} (temporal selection and filtering), and \textbf{DR4} (summarize spatial distributions).
    
\end{itemize}

 \textbf{DR1} was necessary for all tasks as it underpinned the visualization of dengue serotypes in space and time, while \textbf{DR8} served to account for potential bias in the dataset, and to enable hypothesizing future risks, which wsa identified during informal formative evaluations.

%\noindent In addition to these design requirements, two user-oriented goals emerged during development: 

%\begin{itemize} 
%  \item {Ensure ease of use by making GeoDEN intuitive for users, highlighting key features, and enabling rapid insights without requiring in-depth knowledge of the tool's functionalities.}
%  \item {Support independent analysis, separate from GeoDEN, through the capability to download the data in a filtered and processed format.}
%\end{itemize}

%-------------------------------------------------------------------------
\section{GeoDEN}

GeoDEN features three visualisation panels: map, timeline heatmap, and co-occurrence plot (see Figure~\ref{fig:teaser} E, G, and H). Additionally, four control panels facilitate interaction with data processing and display: serotype selector, animation controls, region faceter, and interval length editor (see Figure~\ref{fig:teaser} A, B, C, D, and F).

GeoDEN's dataset contains 3,260 global reports of DENV from 1943 to 2013 \cite{messina_global_2014,messina_global_2014-1}. A supplementary dataset of 289 reports from 2014 to 2020, collected for vaccine deployment and research, is also integrated (J. Messina, personal communication, April 3, 2023) \cite{messina_global_2014}. Each of the 3,549 reports feature a latitude, longitude, country name, year, and four indicator variables which track if each of the four serotypes are reported \cite{messina_global_2014,messina_global_2014-1}. 

Users have the ability to select spatial and temporal contexts, linking all three data visualisations. Spatially, countries are chosen, ranging from one, to a set of countries, to the entire globe. Temporally, consecutive years from 1943 to 2020 can be selected, allowing a span of 1 to 77 years. Users have the ability to facet by serotype, supporting analysis of any combination of serotypes. Faceted serotypes are linked to the Map and Timeline Heatmap visualisations.

Understanding the movements and interactions of serotypes is critical \cite{izmirly_challenges_2020,priyamvada_humoral_2017,houghton-trivino_dengue-yellow_2008} and facilitated by a custom colour scheme, assigning a colour to each serotype to ensure clarity and accessibility. ColorBrewer was initially was used to choose the qualitative four-class colour scheme. However, the only colour-blind friendly option featured two shades of blue and two shades of green, conveying a false relationship between serotypes \cite{harrower_colorbrewerorg_2003,brewer_color_1994}. Instead, we used the site "Coloring for Colorblindness" to generate a palette of four colours that appear different to everyone, supporting DR1, DR4, DR5, and DR7\cite{nichols_coloring_nodate}. Finally, the background colour of the application was selected to be dark grey. This contrasts the four serotype colours and is well-suited for the digital environment.

%------------------------------------------------------------------------
\subsection{Map}

The map panel displays the spatial distribution of dengue reports. Implemented with Leaflet JS (\url{https://leafletjs.com/}), it provides an interactive environment that supports panning and zooming, facilitating analysis at both global and regional scales. Data plotted on the map includes individual reports, regional centroids, trajectories of centroids, and the environmental suitability of dengue.\deleted{s}

\subsubsection{Report Glyphs}

\begin{figure}[htb]
 \centering
 \includegraphics[width=.5\linewidth]{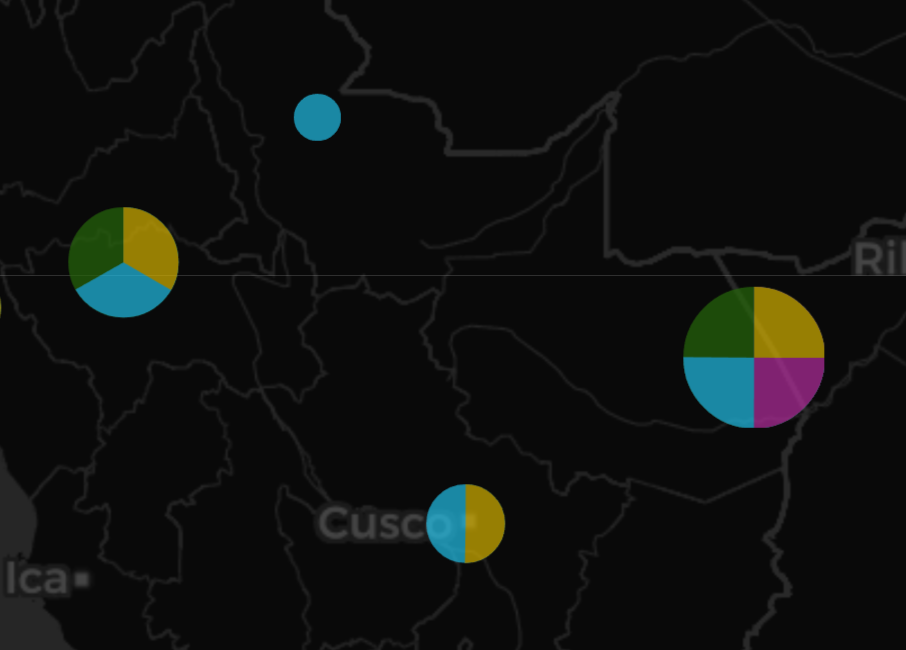}
 \caption{\label{fig:ReportGlyph}
   Report glyphs with different sets of active serotypes.}
\end{figure}

Reports indicate events where a dengue serotype was identified in assays (i.e., tests) in a specific year and location. A report confirms the presence of serotype, but does not specify its prevalence. Consequently, a single report could represent anywhere from one case to 1,000 cases of a specific serotype, for example. Epidemiology experts are accustomed to these types of datasets and their meaning. As seen in Figure~\ref{fig:ReportGlyph}, reports are represented as pie-chart-like glyphs on the map. To \replaced{fulfil}{fulfill} DR1, DR2, and DR3, reports are visualised in time and space for the selected spatiotemporal context. Glyphs are circular and divided into equally sized sections, corresponding to reported serotypes. The area of the glyphs increases with the number of reported serotypes, making co-occurrences more visible (DR7). While we considered both proportional and perceptual scaling of glyphs to the number of serotypes, we decided to instead scale them depending on the 'importance' of that number of serotypes. This is because in the domain, the number of serotypes reported is not conceptualized as a continuous number, but instead an ordinal variable. For our domain experts, each number of serotypes has an independent associated level of risk and importance. Upon hovering, a tooltip labels which serotypes are reported. Upon clicking, a popup provides details about the country, year, and reported serotypes.

We considered using a heatmap or choropleth map instead of individual glyphs. However, these options do not clearly communicate report locations and present difficulties in representing all four serotypes simultaneously, especially given the limitations in the spatial aspect of the dataset. Therefore, individual glyphs were chosen to represent reports, as they effectively convey the place of each report. Glyphs can have diverse appearances \cite{maguire_taxonomy-based_2012}. Our design needed to display the reported serotypes and emphasize serotype co-occurrences. Initially, we considered a glyph-cluster with each glyph resembling a bubble chart, where each reported serotype was represented by a circle, but this created excessive visual clutter. A pie-chart-glyph was presented, and upon further discussion, experts preferred the pie-chart-like glyph. Pie-charts are also already prevalent in visualisations, which reduced the need to train experts on their use \cite{hadfield_nextstrain_2018,arora_serotracker_2021}. The pie-chart-like glyphs scale well to the number of serotypes for any given disease\deleted{ }\added{, al}though they may not scale well to more reports\added{.}\deleted{, in which case other summary views can address occlusion.} \added{In the case of overlapping reports, occlusion is addressed through making the glyphs themselves transparent, as well as the use of other summary views. While a jitter effect could be beneficial, one was not applied to report locations as when looking at reports over time, the transparent nature makes reoccurring reports stand out and appear brighter.} If many \deleted{of }reports occur in a spatial context based on zoom level, proportions of serotypes reported and co-occurrences are not easy to determine through the glyphs alone, which spurred the development of the co-occurrence panel, and partially, the trajectory and cluster views.

%Come Back! Suplimental Material?
% That deleted section is not useless, but is better as supplimentary

\subsubsection{Centroid Glyphs}

\begin{figure}[htb]
 \centering
 \includegraphics[width=.75\linewidth]{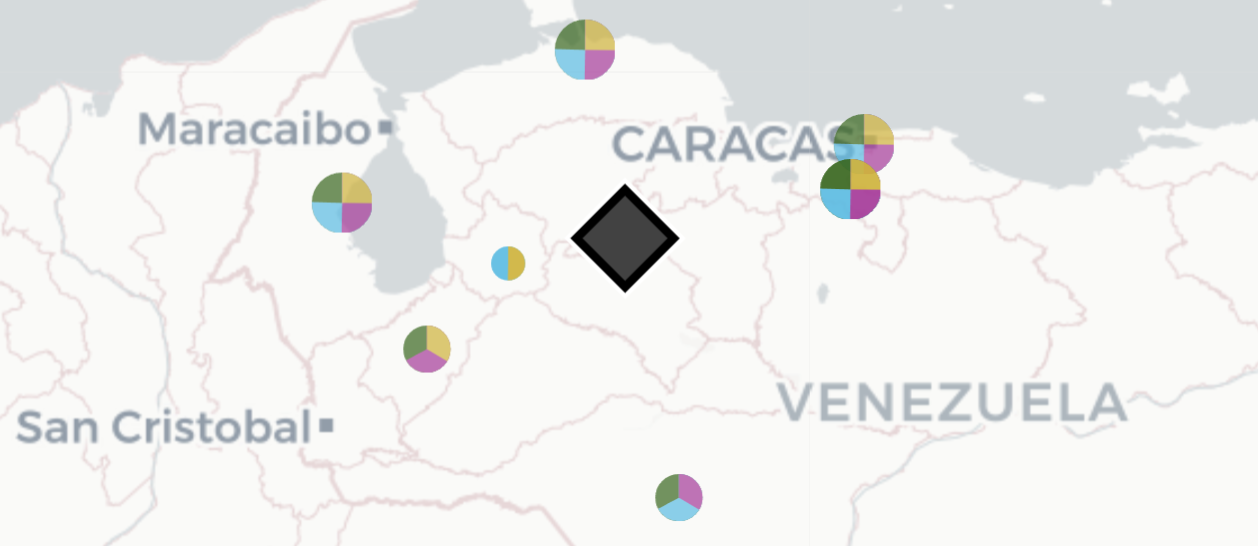}
 \caption{\label{fig:CentroidGlyph}
   A centroid glyph, depicted as a \replaced{grey}{white} diamond, showing the centroid of reports in Venezuela.}
\end{figure}

Following the prototype of the interactive map of reports, our domain experts expressed a need to visualise spatial summary of serotype occurrence (DR4). To meet this requirement, we introduced a glyph representing the average latitude and longitude of reports within the selected spatial region and time frame, seen in Figure~\ref{fig:CentroidGlyph}. GeoDEN, by default, groups reports into continents, but allows users to override using custom regions. Centroids can be calculated for all reports or for each serotype independently, facilitating the analysis of both overall and serotype-specific location summaries. This enables users to compare serotype reporting using only centroid glyphs, offering a more accessible method for analysing summarized spatial occurrences than individual reports. The visualisation of centroids alongside reports provides a complementary approach to spatial analysis.

For the visual representation of centroids, a distinctive point symbol was needed, leading to the selection of a diamond-shaped glyph. The diamond shape offers a clear indication of its center compared to other shapes like squares, triangles, or \replaced{hexagons}{hexigons}. While the option of assigning different shapes to each region was considered, the benefits of a uniform diamond glyph outweighed the advantages of varied shapes. Glyph \replaced{colour}{color} is categorical, representing the region. They range from black to white with two shades of grey. This center \replaced{colour}{color} is used to provide a visual distinction between regions, although this information is also encoded through a popup which shows the name of the region upon hovering or clicking on a centroid glyph. Finally, each glyph is outlined in white and black so that they are easily-visible over any \replaced{colour}{color}. Finally, centroids were found to be useful at both small and large sizes, depending on the user's goal. So, we added the ability to change centroid size on-demand. By default, they are slightly smaller than reports, providing an indication of a group centroid without adding too much visual noise; this is useful when exploring serotype distribution or locating specific reports. However, when comparing groups or serotype distribution within a group, larger centroids were found to be more useful.

\subsubsection{Trajectory Lines}

\begin{figure}[htb]
 \centering
 \includegraphics[width=.75\linewidth]{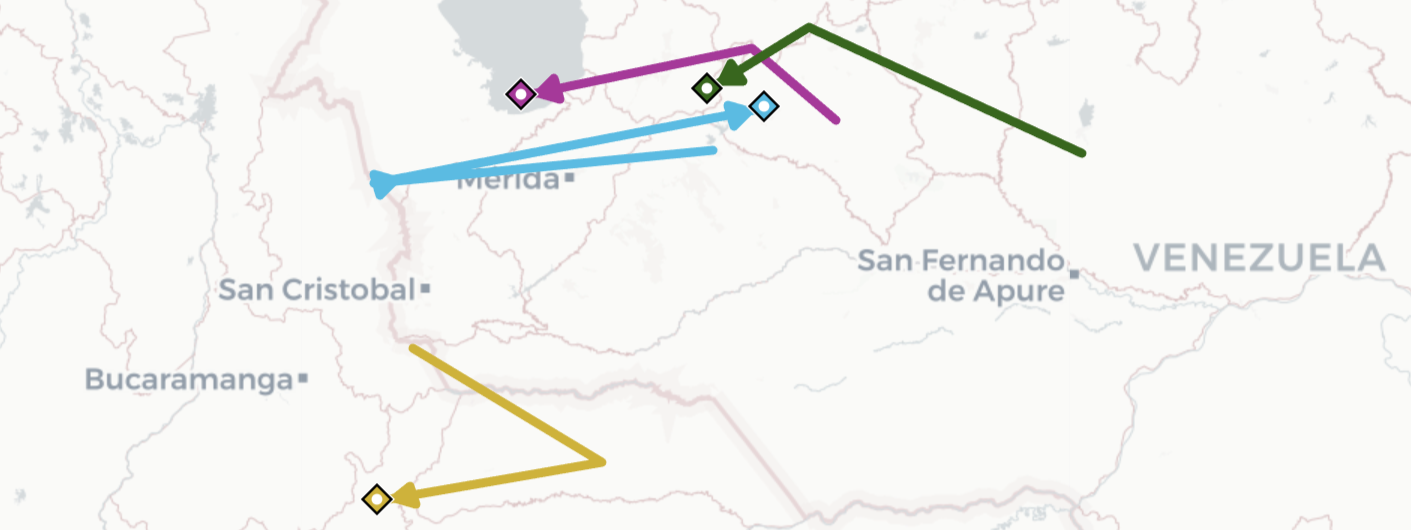}
 \caption{\label{fig:trajectoryLine}
    A trajectory line showing the movement of serotype centroids across Venezuela and Columbia.}
\end{figure}

Our domain experts sought the capability to visualise trajectories: lines tracing the path of centroids over time (DR5). GeoDEN calculates the mean latitude and longitude of reports for each year within the temporal context, and draws a line connecting each year's centroid, as seen in Figure~\ref{fig:trajectoryLine}. Vertices indicate previous years' centroids, and arrows represent the direction of movement. Like centroids, trajectories are calculated by spatial region, and can be calculated for all serotypes or each serotype. While not necessarily representative of actual transmission, trajectory lines provide summary information on where reports tend to be made over time. \deleted{All experts }\added{Experts }were aware of \added{this, yet }\deleted{their limitations but }explicitly requested their implementation.

To visualise trajectory directions, we opted for shape as the primary visual variable instead of colour, width, opacity, or other options, as it proved effective when lines overlap. The use of shape, particularly with the Leaflet Arrowheads package (\url{https://github.com/slutske22/leaflet-arrowheads}), facilitated easy implementation of arrow glyphs. While additional visual variables could be introduced, utilizing only shape proved sufficient to convey the direction of a trajectory. Trajectories do not scale equally across all time scales. When trajectory lines overlap, they become less useful. However, trajectory lines were frequently used in evaluation with domain experts, and were found to be valuable for rapidly discovering new insights.

\subsubsection{Base Map}

\begin{figure}[htp]

\centering
\includegraphics[width=.95\linewidth]{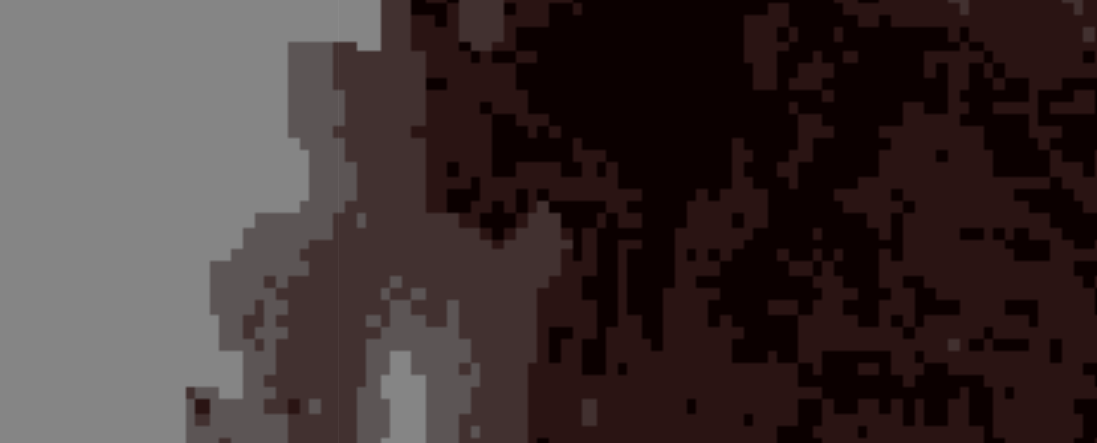}\hfill

 \caption{\label{fig:baseMaps}The environmental suitability \replaced{base map}{basemap}.}

\end{figure}

Dengue occurrences may go unreported, especially given that reports depend on local regulations and research practices\cite{messina_global_2014,messina_global_2014-1,jaenisch_dengue_2014}. Therefore, the expected spatial pattern of reports may not match the observed spatial pattern\deleted{ of reports}. Identifying regions that require further epidemiological research is one of our primary motivations (DR8). To achieve DR8, we used the 2015 environmental suitability map of dengue, generated in Messina et al. 2019\cite{messina_current_2019}. Dengue environmental suitability is a model of the probability of occurrence, calculated using boosted regression trees with covariates that include temperature, precipitation, humidity, and population density. By mapping reports onto a base map of environmental suitability (see Figure~\ref{fig:baseMaps}), we map observed occurrence directly onto expected occurrence. This approach highlights areas with high environmental suitability but no reported cases, prompting users to inquire about the nature of dengue and vulnerability to endemics in those regions, thereby supporting future research and interventions.

\begin{figure}[htb]
 \centering
 \includegraphics[width=.95\linewidth]{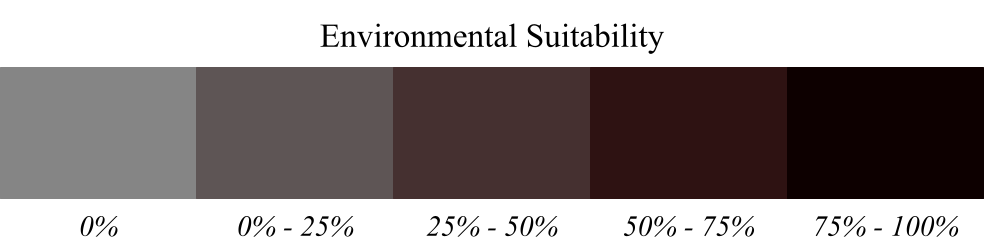}
 \caption{\label{fig:colors}
   \added{Colour scale used to encode environmental suitability.}}
\end{figure}

Environmental suitability is sequential attribute ranging from 0\% to 100\%\added{, as seen in Figure~\ref{fig:colors}}. For our visualization, we separated the data into five equal-interval classes, representing 0\%, 0-25\%, 25-50\%, 50-75\% and 75-100\%. Over these classes, value encodes the most information. Areas more suitable to dengue are darker areas while less suitable areas are lighter. In creating this, we explored several colour scales, including different class breaks, both more and fewer classes, encoding high suitability as lighter values, encoding high suitability as more and less saturated, and even encoding hue for suitability, as well combinations of these. Through this, we found that the class breaks we chose were best, covering large enough sections to be useful and being a relatively intuitive class break. However, to encode that the lowest class is 0, rather than a range, we included saturation in our colour scale. So now, at 0\% suitability there is 0\% saturation, and it increases as suitability approaches 100\%. And including saturation requires a value for hue, and so red was chosen over alternatives as it both signals alarm and contrasts with the four serotype colours.\added{ Finally, we chose to start the colour scale at grey, rather than white, because our report glyph colours were chosen for a dark background.}

%------------------------------------------------------------------------
\subsection{Timeline Heatmap}

\begin{figure}[htb]
 \centering
 \includegraphics[width=.95\linewidth]{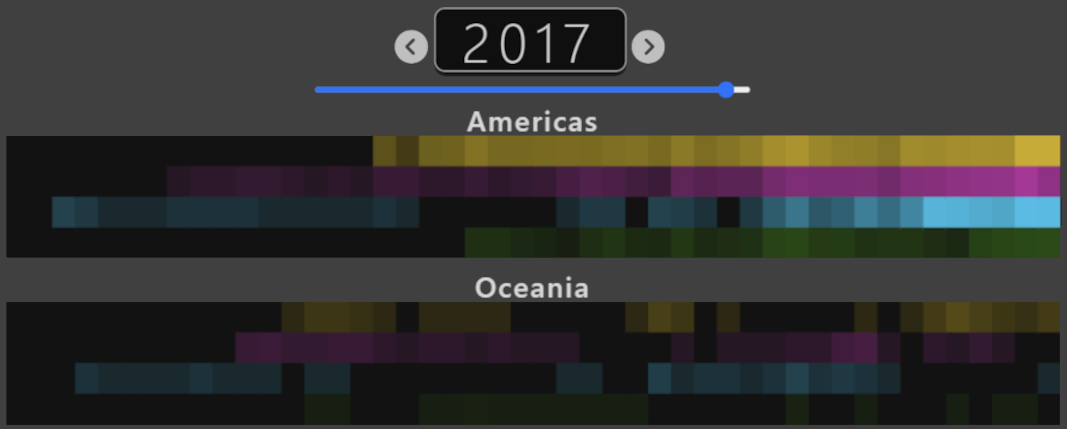}
 \caption{\label{fig:timelineHeatmap}
   The timeline heatmap\deleted{, highlighting reports from 1993 to 1999}.}
\end{figure}

Fulfilling DR1, the timeline view provides another means of comparison and analysis of reports in space and time, and supports temporal selection and filtering (DR3). The timeline, as seen in Figure~\ref{fig:timelineHeatmap}, shows any interval within the full dataset period. Spatial context is reflected through the faceting of timelines by region; each region's events are visualised in their own timeline.

For our experts to make insights about the spread, movement, and co-occurrence of serotypes, occurrence and absence is the most critical information. Using Tableau, we prototyped our timeline view with gantt charts, heatmaps, line charts, streamgraphs, area charts. Heatmaps were \replaced{favoured}{favored} by our experts because they clearly convey occurrence while also encoding frequency. Additionally, heatmaps work especially well when faceted, facilitating the comparison between regions. The heatmap scales well with when visualizing several regions over a large time frame. Details are provided on demand through brushing: when hovering over a cell, a popup relays the year, serotype, and number of reports. Users can also click a cell to see that year, or drag along the heatmap to either extend the active range or change the active year, all of which are brushing techniques that link to both the map and co-occurrence plots.

%------------------------------------------------------------------------
\subsection{Co-Occurrence Plot}

\begin{figure}[htb]
 \centering
 \includegraphics[width=.52\linewidth]{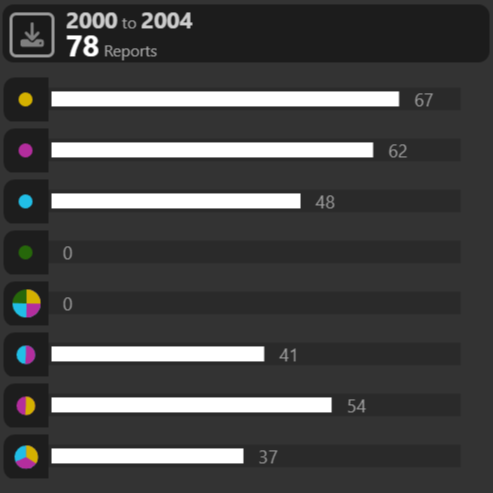}
 \caption{\label{fig:cooccurrencePlot}
   The co-occurrence plot showing proportions of reports from 2000-2004.}
\end{figure}

In the user-defined spatiotemporal context, GeoDEN facilitates the analysis of serotype distribution and co-occurrence in space and time (DR7) through the co-occurrence plot, seen in Figure~\ref{fig:cooccurrencePlot}. This tool enables users to compare the co-occurrence of any serotype combination. User-defined serotype combinations are displayed on the vertical axis, while the frequency of combinations is shown on the horizontal axis.

Understanding patterns in serotype co-occurrence is crucial for the analytical needs of domain users. Therefore, we implemented a visualisation specifically for this purpose. Initially, we prototyped a variation of a node-link diagram, where serotypes are represented by nodes and co-occurrences are encoded through the width of the link. This approach requires faceted graphs to show co-occurrences of three or more serotypes. We also considered a heatmap where serotypes are plotted on the x and y axes, and co-occurrences are visualised at the convergence of the column and row. This approach also requires faceting to show co-occurrences of three or four serotypes. However, a bar chart-like view enables users to compare any desired combination of occurrences without including those that are not of interest. With frequency encoded by the position visual variable, comparisons between combinations are quick. Additionally, this approach scales to any number of serotypes, as any combination can be visualised. The disadvantage of this approach is that users must interact with the view to select serotype combinations. While we considered showing all 15 possible serotype combinations, we found our experts most often compare a small set of serotypes at a time. Therefore, it would introduce visual noise and complexity and jeopardize the targeted analyses our experts perform. Finally, while the data shown is linked to active regions and years, we could imagine implementing a brushing interaction to filter only those combinations in the view. However, this was not a priority, as this effect is possible with the Serotype Selection panel.

%------------------------------------------------------------------------
\subsection{Control Panels}

Users can filter and facet data by serotype, region, and temporal interval. Animation is also included to further allow the discovery of temporal patterns, a feature requested by the domain experts.

\subsubsection{Serotype Selection}

The serotype selection panel enables the selection of which serotypes to visualise and acts as a legend, associating each serotype with its designated colour. Serotype selection assists with DR4, DR5, and DR7. Filtering reduces visual distraction from serotypes not relevant to a given research question. Filtered serotypes are linked to the Map and Timeline Heatmap visualisations.

\subsubsection{Regional Faceting}

\begin{figure}[htp]

\centering
\includegraphics[width=.49\linewidth]{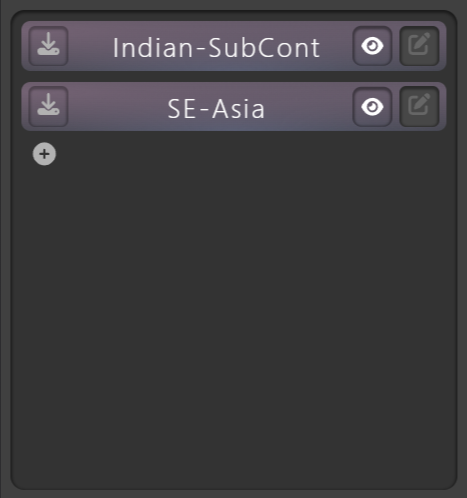}\hfill
\includegraphics[width=.49\linewidth]{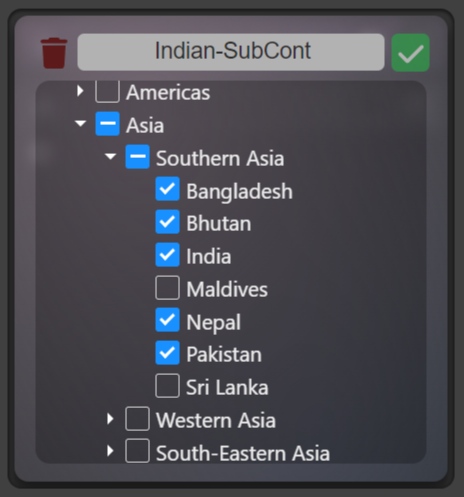}

 \caption{\label{fig:regionFaceting}
    The region faceting panel listing all regions (left) and the region faceting panel editing a custom region (right).}

\end{figure}

The region faceting panel (Figure~\ref{fig:regionFaceting}) serves the purpose of selecting and filtering based on user-defined sets of countries (DR2). Users can make regions visible or invisible, affecting their data representation across all three visualisation panels. Editing regions is facilitated by a tree map, organising countries into three spatial scales: continental, subcontinental, and individual countries. Users define a region though selecting any combination of groups from these scales. All editing and filtering of regions is linked to the Map, Timeline Heatmap, and Co-Occurrence Plot visualisations.

Implementing a clustering algorithm, such as DBSCAN, would automate the definition of regions. However, our expert users preferred the ability to manually define regions, utilizing their expertise. A report’s coordinates often do not reflect the exact location of dengue occurrence \cite{messina_global_2014,messina_global_2014-1}. Consequently, we define regions by country rather than coordinates, despite the fact that this approach does not support analyses at scales smaller than a country.

\subsubsection{Temporal Faceting}

Temporal faceting is enabled by the ability to select an interval of years to view, addressing DR3. The active time period is set by selecting both a current year and an interval length. Various input methods support this, including brushing interaction by clicking and dragging on the timeline, on-screen arrows, arrow keys, and manually typing values. Any change to the interval of years viewed is linked through the Map, Timeline Heatmap, and Co-Occurrence Plot visualisations. This set up was based on requests from domain experts who wanted to ‘trace back’ the history from any given point in time.

\subsubsection{Animation Controls}

An animation has three dynamic variables: the rate of change (time between frames), duration (number of frames an object remains on screen), and order (sequence of frames) \cite{dibiase_animation_1992}. Our interactive system allows the animation to start or stop at any time step, adjust the rate of change via the animation speed, and the duration by changing the interval length. Both on-screen buttons and the spacebar toggle the animation, while an on-screen slider adjusts the animation speed. The chronological nature of the data determines the order of the animation. However, through the temporal controls, users can run the animation in reverse and over small and specific intervals. While this feature does not currently provide a visual history through transition effects, they could be explored in future implementations.

\subsection{Implementation}

All of these features were implemented using the web technologies of JavaScript, HTML, and CSS, and using the packages: D3, JQuery, Leaflet, Leaflet Arrows, TreeJS, PapaParse, and SimpleStatistics. GeoDEN is hosted at \URL{http://GeoDEN.net}, and its source code and data is made available on \URL{https://github.com/geohai/GeoDEN}.

%------------------------------------------------------------------------
\section{Evaluation}

Our three domain experts in dengue and epidemiology each participated in two complementary evaluations: an insight-based evaluation and a value-driven evaluation. The insight-based evaluation is our primary approach, ideal for testing GeoDEN's ability to facilitate insight generation—insight generation being GeoDEN's purpose to our expert users \cite{north_toward_2006}. To test GeoDEN's value more holistically, we also conducted a value-driven evaluation. Combined, these evaluations seek to identify the successes and failures of our methodologies in the design and development of GeoDEN.

\subsection{Insight-Based Evaluation}

Each insight-based evaluation session was performed over an hour-long period, and included a think-aloud approach, as seen in Protocol Analysis \cite{fan_vista_2019,arias-hernandez_pair_2011}. Think-aloud approaches involve the domain expert verbally stating what they are thinking and trying to achieve with the tool.
Because participants may forget to iterate their thoughts, using a think-aloud approach alone has the possibility of participants forgetting to iterate their thoughts, we employ a constant \added{dialogue }\deleted{dialog }between the epidemiology and visualisation expert, as seen in Pair Analytics \cite{arias-hernandez_pair_2011,hnatyshyn_molsieve_2024}. 
This conversation clarifies what the domain expert is thinking and doing during the evaluation.
In each session, the domain expert was asked to use GeoDEN to verify their own knowledge of dengue serotype patterns and look for novel insights.  After, they were asked to provide feedback on their experience and thoughts on potential further development and use cases.

\subsubsection{Expert 1}

\noindent \textbf{Insights Validated: }
Expert 1 used GeoDEN’s map panel to confirm that dengue was initially confined to Southeast Asia, iterating year by year from 1943. Moving forward in time, they successfully verified the emergence of serotype 3 in Puerto Rico in 1963 and its subsequent spread throughout the Americas, as well as a significant burst of serotype 1 reports in 1977. Both findings were derived from careful observation of the timeline and map panels. Expert one continued to confirm the occurrence of pivotal events, including the dengue epidemic in Cuba in 1981 and research emerging from Peru’s Iquitos research center in the 1990s. Using the co-occurrence plot alongside the map, they validated the appearance of severe cases, noting the first recorded instances of Dengue Hemorrhagic Fever (DHF) in the Americas in 1981. These were consistent with severe manifestations already prevalent in Asia, reinforcing the connection between serotype interactions and disease severity \cite{ohainle_dynamics_2011,priyamvada_humoral_2017,houghton-trivino_dengue-yellow_2008}. The map panel also allowed the expert to confirm the hyperendemic nature of Southeast Asia, where all four serotypes coexisted, and to observe shifts in serotype dominance across different locations and times. By tracking the centroids of serotype-specific trajectories, they verified dominant serotype circulation in West Africa between 2006 and 2010. Similarly, the expert tracked movement between southern India and Sri Lanka, verifying that serotypes in the area were consistently present and co-reported over time.

\noindent \textbf{Insights Generated: }
In the course of their exploration, the expert generated several novel insights. While analyzing West Africa, they noticed an unexpected large movement of serotype 2 between Nigeria and Senegal. Doubting its credibility based on their domain knowledge, they divided West Africa into two smaller regions—eastern and western. This adjustment revealed that the previously perceived single transmission bubble in the region was likely two separate zones. Another discovery emerged during their investigation of the Indian subcontinent. The expert observed a consistent movement of all dengue serotypes between southern India and Sri Lanka. This prompted them to hypothesize about the synchronicity of serotype movement. Using the co-occurrence plot, they generated bar charts of serotype combinations, confirming that all serotypes tended to be reported simultaneously. They further refined their analysis by narrowing the time window to three years and noticed a disconnect between northern and southern India, but noted unreliable patterns due to limited data resolution. The expert also explored the possibility that Bangkok acts as a "pacemaker" for dengue transmission across Southeast Asia\cite{cummings_travelling_2004}, testing this hypothesis using centroid trajectories. They observed the evolution of research and reporting activity across the region, noting that while reports in the 1970s were concentrated in Myanmar and Bangkok, Vietnam saw a surge in activity by 1995. Post-2000, multiple countries established reporting systems in parallel, underscoring the evolving landscape of research in Southeast Asia. Increased activity in Myanmar and Vietnam over time further clarified the influence of research hubs on the dataset and informed hypotheses about the data’s temporal and spatial coverage.

\noindent \textbf{Feedback: }
Expert 1 praised the map panel for clearly depicting spatiotemporal patterns at various scales and the co-occurrence plot for uncovering serotype interactions. Centroid trajectories effectively \replaced{modelled}{modeled} transmission dynamics, though they highlighted spatial resolution limits. Despite these limits, expert 1 commented that “it’s beautiful how you can see how things populate over time; how the Africa continent is rather quiet.” Many periods and regions do not show many reports despite likely having dengue, however, expert 1 finds that this “is not a weakness, [rather] a good description”.

\subsubsection{Expert 2}

\noindent \textbf{Insights Validated: }
Expert 2 began in Africa, interested in validating their knowledge of the disconnect between dengue occurrence and reports. Filtering the data to only show reports from Africa over 10 year intervals, they animated through the whole timeline, quickly finding what they expected: that reporting in Africa was broadly sparse, especially areas one would expect dengue reports given high environmental suitability.
Next, expert 2 changed their attention to the Americas, filtering data to only show reports from the Americas. Continuing to visualise reports in 10 year intervals, they saw all four serotypes eventually spread all around the Americas, and observed that regardless of which serotype arrived first in a country, all four serotypes eventually spread throughout the region. This finding aligned with their prior knowledge but underscored its significance for vaccine development, as targeting a single serotype may have long-term implications for immunity and disease severity.

\noindent \textbf{Insights Generated: }

Again filtering to just reports from Africa, expert 2 was interested in discovering new things its sparse reporting. Expert 2 noted how the centroid moved when animating through all data. So, they tried again with serotype-specific centroids, ad between these centroids and the co-occurrence plot, they found that serotype 2 moved all around Africa and was both the most reported and mobile of the serotypes. They noted this pattern could be the result of several factors including which tests were administered, how they were administered, difference in symptoms that arose from the serotypes, exposure to multiple serotypes, or others. Going deeper, expert 2 filtered to only see reports from central Africa, a region which has few reports despite high environmental suitability. Serotype 2 was again the most prevalent with sparse reports of serotypes 1 and 3 and no reports of serotype 4. This revealed to them that research should be done in the Democratic Republic of Congo and other nearby countries as the expected spatial patterns of occurrence did not match reported spatial patterns of occurrence. Testing would help to determine whether dengue is truly absent or underreported.
Next, expert 2 looked at the earliest reports of dengue in the 1940s, revealing that serotype 1 was reported in Japan in 1943 and Hawaii in 1944, leading to the hypothesis that transmission occurred between these locations. This new insight invites further research into whether dengue spread directly from Japan to Hawaii or through intermediaries. Finally, using the timeline heatmap, they tracked the rise and fall of dengue reports in the Americas, noting that preventative control policies in the 1980s and 1990s likely led to decreased reports, while relaxed measures in the 2000s coincided with fluctuating reports.

\noindent \textbf{Feedback: }
Overall, expert 2 found GeoDEN intuitive and was successful in both verifying known insights as well as discovering new ones. They mentioned that “watching it as movement over time, rather than as a static image, really does… make you ask a lot more questions [about] how things spread”; “being able to decide the interval... you start to see new patterns that you don’t get with [static maps]”. They also mentioned that “[GeoDEN] does need to be used by people who understand a basic level of epidemiology”, given the biases of the dataset.

\subsubsection{Expert 3}

\noindent \textbf{Insights Verified: }
Expert 3 aimed to verify their knowledge of serotype patterns in Africa: that it originated in West Africa and that it is sparsely reported. They used the region faceting panel to select and show data from Africa as one region, the interval length panel to show 10 years of data, and the timeline to select the earliest report, validating its start in west Africa. Using the environmental-suitability \replaced{base map}{basemap}, they noted it started in a high-suitability area. The lack of reports confirmed a need for more data collection in Africa, supported by their knowledge that, in Africa, Dengue is often misdiagnosed as malaria.
Connected to their work, Expert 3 decided to analyse the trajectories of serotypes in Brazil, with a focus on higher spatiotemporal resolutions. They filtered data to only show reports from Brazil between 1980 and 2020. Focusing on the noticeability of each report, they switched to the dark \replaced{base map}{basemap}. Finally, they adjusted the centroids to trace paths of each serotype. With this, they verified serotype 3's introduction and spread in the 2000s. However, they were unable to analyse two patterns the way they wanted to due to data limitations: the movement of serotype 4 over the course of weeks, and the occurrences of serotypes specifically in the city of Rio de Janeiro. This is a limitation, as the underlying dataset does not have that level of spatial or temporal resolution, so GeoDEN cannot facilitate that level of analysis.

\noindent \textbf{Insights Generated: }
Starting in Africa and using the map and timeline panels, Expert 3 quickly discovered that serotype 4 was the least reported serotype in Africa, only appearing in 1983 and 1995. Next, expert 3 examined how PCR diagnostic tests from the 1980s affected global dengue reports. They toggled all regions and serotypes to be active, set the current year to 1945, and used 10 year intervals. Iterating forward through time, they made note of the fact that before 1980, most reports were from Asia—a pattern they were not aware of. They also noticed reports greatly increase from the 70's to 80's in Asia. They verified this using the co-occurrence panel. In 1970-1980, there were 242 reports while in 1980-1990, there were 541; more than than double.
Expanding out to a global view, expert 3 observed the shift of reports from serotype 1 in the 1980s to serotypes 2 and 3 by the 2000's. They discovered that serotype 3 was reported the most in the early 2000s, especially in 2002. They also found that both the Americas and Oceania have a similar pattern of reports being introduced: first serotype 3, then 2, then 1, and finally 4. This could be due to available tests, a coincidence, or a pattern in how serotypes interact at the scale of populations.

\noindent \textbf{Feedback: }
Expert 3 commented that GeoDEN was intuitive and easy to use, noting they had a positive experience. While they were able to verify insights and generate new ones, they wanted to see the tool grow to support datasets at finer spatiotemporal scales, as well as datasets input by user.

\subsection{Value-Driven Evaluation}

The value-driven evaluation, proposed by Stasko in 2014, is a qualitative approach to estimate the value of a visualisation\cite{stasko_value-driven_2014}. Stasko's methodology evaluates a visualisation along four metrics: its ability to minimize \textbf{time} needed to answer a variety of questions, its ability to facilitate the discovery of \textbf{insights} and spur insightful questions, its ability to convey the \textbf{essence} of the data, and its ability to generate \textbf{confidence} about the data and its domain \cite{stasko_value-driven_2014}. This method was further developed by Wall et al. in 2018 to provide a quantitative evaluation \cite{stasko_value-driven_2014, wall_heuristic_2019,wang_emotional_2019,deng_visualizing_2024,ceneda_heuristic_2023}. Wall et al. first identify a set of guidelines meant to encapsulate each metric, and then further break these into heuristics. Each heuristic is rated on a scale of 1-10 by visualisation experts \cite{wall_heuristic_2019}.
Without a panel of visualisation experts, we opted to use Stasko's qualitative evaluation. However, to better to elicit the value of GeoDEN, we ask our experts how well it satisfies each of the 10 guidelines developed by Wall at al.The full results are included in supplementary materials.

    \subsubsection{Insight}
    \begin{enumerate}
        \item \emph{A visualization’s ability to spur and discover insights and/or insightful questions about the data.} To this prompt, the experts were positive. Expert 1 mentioned that it helped spur new questions expert 3 was happy with the spatiotemporal exploration capabilities, however found summaries for certain regions to be limited.
        \item \emph{Does the visualization facilitate answering questions about the data?} All three experts said "yes", with expert 2 highlighting use of the environmental suitability map while expert 1 highlighted GeoDEN's ability to understand movement over time.
        \item \emph{Does the visualization provide a new or better understanding of the data?} All three experts affirm this, but to varying degrees. Expert 1 mentions that it makes aspects more visible, especially ones already known; expert 2 mentions yes, especially temporally; and expert 3 says it is the only understanding for this type of data.
        \item \emph{Does the visualization provide opportunities for serendipitous discoveries?} Expert 1 was adamant about its ability for find new connections and discoveries. Expert 2 also agreed, but distinguished between discovering outbreaks of serotypes at regional scales and larger spatiotemporal patterns at global scales.
    \end{enumerate}
    Our experts found it certainly facilitates answering questions about the data, provides a new or better understanding of the data, and provide\added{s} opportunities \replaced{for}{to} serendipitous discoveries. Overall, responses to these prompts were positive with just a couple ambiguous responses.

    \subsubsection{Confidence}
    \begin{enumerate}[start = 5]
        \item \emph{A visualization’s ability to generate confidence, knowledge, and trust about the data, its domain and context.} Experts did not say it supported high confidence with expert 3 mentioning that because the data is only reports, it has bias, and expert 1 mentioning that all hypotheses need to be validated with external information.
        \item \emph{Does the visualization help to avoid making incorrect inferences?} Expert 1 again mentions that questions and hypotheses need to be validated; expert 2 mentioned that there is reporting bias, which can be misleading, but users of the visualisation will be aware of these limitations; expert 3 said it depends on the user who made the inference.
        \item \emph{Does the visualization help to understand data quality?} Experts 1 and 2 say yes, however expert 3 says 'not really' as the concept of serotypes must be known.
    \end{enumerate}
    Experts agreed that GeoDEN does not help users to avoid making incorrect inferences, although each had their own take on it, mentioning inherit reporting bias, the user, and validating insights externally. Experts were split on whether it helps users understand data quality, with experts 1 and 2 thinking it does, but expert 3 not agreeing. Overall, the confidence of insights generated is not high, and requires an understanding case reports as well as the external validation of insights.

    \subsubsection{Essence}
    \begin{enumerate}[start = 8]
        \item \emph{A visualization’s ability to convey an overall essence or take-away sense of the data.} This general prompt was responded to positively with one expert mentioning that it \replaced{distils}{distills} the data into a cohesive message, although that the visualization can be overwhelming, showing so many different properties of the data at once. They also mentioned that it does not provide info on which part of the data is more robust or essential than others.
        \item \emph{Does the visualization provide a big picture perspective of the data?} All three experts affirm this, with expert 1 saying it captures different aspects than other data.
        \item \emph{Does the visualization provide an understanding of the data beyond individual reports?} All three experts say "yes", with expert 1 mentioning that it provides a different perspective on the data.
    \end{enumerate}
    The visualization was both found to certainly provide an understanding of the data beyond individual reports and a 'big picture' of the data in general. Expert 1 did note, however, that GeoDEN does not encode which parts of the data are more robust.

    \subsubsection{Time}
    \begin{enumerate}[start = 11]
        \item \emph{A visualization’s ability to minimize the total time needed to answer a wide variety of questions about the data.} While it was noted that the visualisation had a high ability to generate insights quickly, expert 1 highlighted that 'time' is not important here, rather it is about generating different insights than they otherwise could have.
        \item \emph{Does the visualization support the understanding of multiple pieces of information at once?} All three say "yes", with expert 1 highlighting that it is holistic and expert 2 highlighting that GeoDEN makes it easy to compare different places and times.
        \item \emph{Does the visualization provide mechanisms for quickly seeking specific information?} Experts 2 and 3 both say "yes", one highlighting the importance of the use case and the other abilities of filtering data to a specific time and location. Expert 1, however, again made it clear they found this question to miss the mark, instead saying that it is not about saving time but instead generating insights than they otherwise could have.
    \end{enumerate}
    All experts both found that GeoDEN supports understanding multiple pieces of information at once and provides mechanisms for quickly seeking specific information. However, expert 1 made it clear that they found this section less important; they say GeoDEN supports generating insights that otherwise would not have been, and that this is its true value. We include this to highlight that expert 1 holds insight to be more important than other metics of value.
    
    \subsubsection{Summary}
    The ability for GeoDEN to drive insights and discovery appears to be the most valuable aspect for our expert users. They answered positively to these questions, and noted it when responding to the efficiency of the visualisation that time is not of importance when it helps make insights that otherwise wouldn't have been discovered. GeoDEN was also found to be effective at showing the essence of the data as well as reducing time to make discoveries. However, where GeoDEN failed was in its ability to show the confidence of the data. The experts did not critique GeoDEN, given that the underlying dataset lacks certainty indicators, making it difficult to determine the credibility of individual reports. A dataset of case reports has certain benefits and drawbacks—the benefits being that they can cover a wide spatiotemporal range, the drawbacks being that reports do not properly represent prevalence and have a high degree of uncertainty, as some serotypes may not have been tested.
    GeoDEN is a valuable addition to visual analytics for epidemiology given its efficacy in generating insights, its efficiency, and its ability to convey the essence of the dataset. However, its value is limited by the confidence in the insights generated.

%------------------------------------------------------------------------
\section{Discussion}

GeoDEN demonstrated effectiveness in enabling epidemiology experts to verify and discover new insights at both global and regional scales. Even though it is the most comprehensive dataset on dengue available, analyses were primarily limited by the temporal span and spatiotemporal resolution of the dataset, rather than GeoDEN.

\subsection{Evaluation of Design Requirement Implementation}

\noindent \textbf{DR1} - Visualisation of dengue serotype reports in both time and space proved effective. The map panel predominantly facilitated the exploration of individual reports and small-scale events due to its clarity in presenting filtered data. However, more abstract conclusions, such as the timing of serotype introductions and the proportion of serotype reports, were often derived from the co-occurrence and timeline panels.

\noindent \textbf{DR2} - Our design allowed for the selection and filtering of geographic data, providing details on the number and timing of reports. Users consistently utilized the region selection and faceting panel to choose specific geographic regions for in-depth report analysis. Notably, expert 3 identified hyperendemic regions and potential bifurcation in transmission bubbles using this feature. However, expert 3 expressed a desire for the ability to draw custom regions on the map, allowing for the selection of complex shapes using lasso selection beyond the limitation of country boundaries.

\noindent \textbf{DR3} - Time selection and filtering, achieved by editing the active year and interval length, effectively enabled temporal analysis. No problems with this system were found and all three experts used both features greatly and commented on their ease of use.

\noindent \textbf{DR4} - Serotype selection and filtering facilitated comparative analysis for experts, yielding insights into serotype introduction and frequency. While experts found it straightforward to navigate, it was not frequently used, especially when compared to the selection and filtering of geography or time.

\noindent \textbf{DR5} - Trajectories of serotypes were analysed and compared frequently by experts, leading to many insights around the movements and transmission of serotypes.

\noindent \textbf{DR6} - All domain experts frequently utilized the animation panel in conjunction with other temporal selection tools. Animation played a crucial role in discovering transmission events and providing a fresh perspective on the data, prompting new questions. Expert 2 specifically praised its intuitiveness and discovered novel insights, such as the potential transmission of serotype 1 from Japan to Hawaii.

\noindent \textbf{DR7} - Co-occurrences between serotypes were visualised and analysed by all domain experts using all visualisations, enabling inferences about interactions between serotypes that could be associated with severity. Expert 3, whose research focuses on severity arising from cross-reactivity, expressed satisfaction with the inclusion of this feature in the final implementation.

\noindent \textbf{DR8} - The ability to find areas where expected spatial patterns differ from observed spatial patterns was straightforward with the environmental suitability \replaced{base map}{basemap}, which directly contrasted the plotted location of reports. This analysis was even described as beautiful for visualising the lack of data in Africa, which all three domain experts quickly identified and investigated.

\subsection{Limitations}

While GeoDEN proved to be an effective and valuable tool, it is not perfect. It features a heavy reliance on users being experts and already knowing contextual information, and awareness of the limitations in data, such as how reports suffer from external bias related to funding and awareness. This issue revealed itself in a disconnect between expert 1 and expert 3 where expert 1 found a novel insight that dengue was primarily reported in Asia for many years; expert 3 already knew of this pattern, and that it was particularly in southeast Asia. A novel insight for one researcher was a verified insight for another. If contextual events and information were stored and communicated through GeoDEN, it would allow all researchers to better understand contextual events without the need to actually know each one independently, and would therefore centralize knowledge. With the advent of Large Language Models (LLMs), enabling quick summarization of research literature, further textual enhancement of the tool can be envisioned.

Further, the underlying dataset has inherent uncertainty that is not visualised. A report’s latitude and longitude does not necessarily represent the specific location of observation: a report can represent a country, a subnational region, a city, or a specific location. However, full details on what any given report represents was not a part of the original published dataset. This means we, as visual analytics researchers, do not know if a given point refers to a specific location or a whole subnational region. 
While we chose not to encode this information, there are several ways the design could incorporate the level of detail of a case report. For example, the glyph design could be modified so that reports at the country level are represented with a white outline, while more specific reports include a white dot at the center. Size and opacity could also encode the level of detail: country-level points could be larger, more transparent, and have edges that fade to full transparency, while more specific reports could use smaller, more opaque glyphs with hard edges. One final encoding we can imagine would be to instead visualise reports through a choropleth map, where the level of specificity determines how an area is shaded from country to more specific levels, and could change how shaded an area is. While this would introduce other challenges, it would not overrepresent the specificity of a report's location, possibly being the best solution for a project with similar data but different tasks.
Further, the original dataset only extended from 1943 to 2013, with a supplemental dataset filling in some data from 2014 to 2020; this restricts the insights that can be gleaned from 2013 to 2020. Despite these limitations, our experts are used to working with such datasets and therefore, the value of GeoDEN was not impacted beyond a reduced confidence in insights generated.

\subsection{Scalability and Generalizability}
We discussed the scalability of individual visualisations in each section above. As for generalizability, our methods for developing a VA tool tailored to solve the epidemiological problem of visualising the movement and co-occurrence of serotypes can be applied to other disease incidence datasets with serotypes. One requested implementation would expand upon GeoDEN to allow higher resolution spatiotemporal datasets, so that the same analyses might be performed over days across a city rather than over years across countries. Additionally, our combination of insight-based evaluation and value-driven evaluation can be used by future VA researchers. Using both provides a comprehensive analysis of how well the tool generates new insights and hypotheses while also providing a systematic heuristic evaluation of the system's value. This evaluation approach is suited for tools that feature iterative user design with expert users, as it is both cost effective and impactful.

\subsection{Lessons Learned from Collaborative Design Process}
Collaborating closely with epidemiology experts has highlighted the specificity of design requirements and that sometimes more sophisticated algorithmic approaches are not received by the users. For instance, we recommended automated algorithms such as DBSCAN for finding clusters, which was not welcomed by users, as they required more manual control and understanding given the uncertainty in the data. The perceived value of the tool was high, especially in terms of insight generation. This is due to our strong focus on addressing experts' design requirements, which were largely about enabling insights. Although the dataset's uncertainty reduced confidence in the tool, all users found it a significant step forward in generating insights and making sense of the data over time and space. Lastly, and unsurprisingly, we found frequent meetings with users during the formative stage to be helpful in correcting paths towards meeting user and use requirements.
%------------------------------------------------------------------------
\section{Conclusion and Future Research}

In this article, we presented the design of GeoDEN, a visual \added{exploration }\deleted{analytics }tool for\added{ the} exploratory and confirmatory analysis of\added{ dengue serotype movement}\deleted{ epidemiological data}, and both insight-based and value-driven evaluations \deleted{of the tool }using a long-term and global dataset of dengue serotype reports. Insights are enabled by our three linked visualisations and supported by animating the selection and faceting of data in the spatial, temporal, and attribute dimensions. Our three domain experts confirmed and refined their existing hypotheses, as well as discovered novel insights warranting further investigation by the epidemiology community. GeoDEN enables epidemiologists to understand serotype movements and interactions at the global and regional scales, as well as inform aid and support intervention, including vaccination campaigns. The visualisations and systems designed for GeoDEN offer a versatile framework that can extend to other serotype report datasets at a range of resolutions.

One potential extension is to include molecular epidemiology data into the dataset, which would enable researchers to draw conclusions about suspected movement patterns with more certainty if viruses are phylogenetically closely related. However, this requires substantial data collection, involving the extraction of this data from the original research articles used in the creation of the dataset. This has the potential for automation using Large Language Model (LLM) summarization techniques. Related to this, incorporating and visualising additional geolocated textual information gleaned from research articles can help further contextualize the report data. Techniques such as geospatial word cloud visualisation or topic mapping, as seen in Interactive Learning for Identifying Relevant Tweets to Support Real-time Situational Awareness, can be used to further enhance the analytical capabilities of GeoDEN, supercharged by the strong capabilities of LLMs and generative summarization techniques \cite{snyder_interactive_2019}. Finally, we would implement the ability to draw regions using lasso polygons, provided a dataset with more certainty of occurrence locations.

\section*{Acknowledgements}
This work was supported by the U.S. National Science Foundation under Grant 2026962, and University of Colorado Boulder RIO Seed Grant.

\section*{Data Availability}
Code and data used are available at  \URL{https://github.com/geohai/GeoDEN}. Original datasets are available at \URL{https://figshare.com/s/d7d7871d00afe2870619} \cite{messina_current_2019,messina_global_2014,messina_global_2014-1}. 

\section*{Conflict of Interest}
We have no conflicts of interest to disclose.
%------------------------------------------------------------------------

\bibliographystyle{eg-alpha-doi} 
%\bibliography{egbibsample}
\bibliography{main}

\end{document}